\documentclass[twocolumn,superscriptaddress,aps,pra]{revtex4}
\usepackage{graphicx,amsmath,amssymb}
\usepackage{dcolumn,fancyhdr}
\usepackage{bm}
\usepackage{times}
\usepackage{txfonts}
\usepackage{epstopdf}
\usepackage{latexsym}
\usepackage{epsfig}

\newcommand{\ket}[1]{\left\vert#1\right\rangle}

\newcommand{\bra}[1]{\left\langle#1\right\vert}

\usepackage{color}
\definecolor{Blue}{rgb}{0,0,1}
\definecolor{Red}{rgb}{1,0,0}
\definecolor{Green}{rgb}{0,1,0}
\definecolor{Purp}{rgb}{.2,0,.2}
\definecolor{white}{rgb}{1,1,1}

\usepackage{graphicx}
\usepackage{bm}
\usepackage{epstopdf}
\usepackage{latexsym}
\usepackage{epsfig}

\begin{document}
\title{Global quantum correlations in finite-size spin chains}

\author{S. Campbell}
\affiliation{Quantum Systems Unit, OIST Graduate University, 1919-1 Tancha, Onna-son, Okinawa 904-0495, Japan}
\affiliation{Department of Physics, University College Cork, Republic of Ireland}

\author{L. Mazzola}
\affiliation{Centre for Theoretical Atomic, Molecular, and Optical Physics, School of Mathematics and Physics, Queen's University, Belfast BT7 1NN, United Kingdom}

\author{G. De Chiara}
\affiliation{Centre for Theoretical Atomic, Molecular, and Optical Physics, School of Mathematics and Physics, Queen's University, Belfast BT7 1NN, United Kingdom}

\author{T. J. G. Apollaro}
\affiliation{Dipartimento di  Fisica, Universit\`a della Calabria,
87036 Arcavacata di Rende (CS), Italy} \affiliation{INFN - Gruppo
collegato di Cosenza}
\affiliation{Centre for Theoretical Atomic, Molecular, and Optical Physics, School of Mathematics and Physics, Queen's University, Belfast BT7 1NN, United Kingdom}

\author{F. Plastina}
\affiliation{Dipartimento di  Fisica, Universit\`a della Calabria,
87036 Arcavacata di Rende (CS), Italy} \affiliation{INFN - Gruppo
collegato di Cosenza}

\author{Th. Busch}
\affiliation{Quantum Systems Unit, OIST Graduate University, 1919-1 Tancha, Onna-son, Okinawa 904-0495, Japan}
\affiliation{Department of Physics, University College Cork, Republic of Ireland}

\author{M. Paternostro}
\affiliation{Centre for Theoretical Atomic, Molecular, and Optical Physics,
School of Mathematics and Physics, Queen's University, Belfast BT7 1NN, United Kingdom}

\begin{abstract}
We perform an extensive study of the properties of global quantum
correlations in finite-size one-dimensional quantum spin models at
finite temperature. By adopting a recently proposed measure for
global quantum correlations [C. C. Rulli, and M. S. Sarandy, Phys.
Rev. A {\bf 84}, 042109 (2011)], called {\it global discord}, we
show that critical points can be neatly detected even for
many-body systems that are not in their ground state. We consider
the transverse Ising model, the cluster-Ising model where
three-body couplings compete with an Ising-like interaction, and
the nearest-neighbor XX Hamiltonian in transverse magnetic field.
These models embody our canonical examples showing the sensitivity
of global quantum discord close to criticality. For the Ising
model, we find a universal scaling of global discord with the
critical exponents pertaining to the Ising universality class.

\end{abstract}
\date{\today}
\maketitle


Entanglement and criticality in quantum many-body systems have
been shown to be  strongly and intimately connected
concepts~\cite{Osterloh,osborne}. The body of work performed with
the aim of grasping the implications that critical changes in the
ground state of a given Hamiltonian model have for the sharing of
entanglement by the parties of a quantum many-body systems is now
quite substantial~\cite{Amico}. This has resulted in important
progresses made in our understanding of the interplay between
critical phenomena of interacting many-body systems and the
setting up of genuinely quantum features. In turn, such success
has proven the effectiveness of the cross-fertilization of quantum
statistical mechanics by techniques and interpretations that are
typical of quantum information theory.

Yet, it has recently emerged that the way correlations of
non-classical nature manifest themselves is not necessarily
coincident with entanglement, and a much broader definition of
quantum correlations should be given~\cite{Ollivier,Henderson}.
This is encompassed very effectively in the formulation of
so-called quantum discord as a measure striving to capture the
above-mentioned broadness of quantum correlations~\cite{Modi}. In
analogy with the case of entanglement, the relation between
quantum discord and the features of quantum many-body models is
fundamentally interesting for the understanding of the role that
the settlement of quantumness of correlations play in determining
the critical properties of such models. A systematic analysis in this sense, which has only recently been considered~\cite{varie,giorgi,werlang,tomasello,campbell2,sarandy1,gedik}, is thus highly desirable. This is even more important given that some of the
investigations performed so far have indicated that quantum
discord is more sensible than entanglement in revealing quantum
critical points~\cite{werlang}, even for systems that are not at zero
temperature~\cite{sarandy2}.

This is a particularly relevant result, whose validity should also
be checked for models that are both finite sized and at finite
temperature. The motivations for such an endeavor stem from the
fact that, likely, the properties of quantum many-body systems
will be addressed experimentally in systems consisting of, for
instance, cold atoms loaded in optical potentials or trapped ions,
as in very recent ground-breaking
experiments~\cite{Simon,Friedenauer,Lanyon,Britton,Islam}. At
variance, the studies performed so far have mostly dealt with
systems at the thermodynamic limit. Moreover, intuitively, one
would expect global measures of general quantum correlations to be
well suited to reveal the subtle features at hand here, given that
some of the critical changes occurring in the lowest-energy state
of many-body systems truly involve (quasi-)long-range influences
among the parties. Such an analysis is made very difficult, both
at a theoretical and computational levels, by the lack of
unambiguous measures of multipartite entanglement in mixed states.

In this paper, we study the relation between criticality and
global quantum correlations in finite-size systems at non-zero
temperature by using a measure of global quantum correlations
recently put forward in Ref.~\cite{RulliArX11} and employed by
some of us in~\cite{campbell2} for a quantum many-body system at
zero temperature. As canonical examples, we study one-dimensional
models that are of genuine physical interest due to the
non-trivial features of their phase diagrams, such as the
transverse-field Ising model, the open-boundary XX model in
transverse magnetic field~\cite{Katsura}, and the so-called
cluster-Ising model introduced in Ref.~\cite{Son}. The latter
interpolates between the standard antiferromagnetic Ising
Hamiltonian and a topologically ordered {\it cluster} phase. Our
study shows the ability of global discord to detect critical
points. Moreover, for specific cases among the examples addressed
in our work, we bring evidence of a finite-size scaling for global
discord and its derivative that are closely related to the
behavior of macroscopic features such as the magnetization.

The rest of this manuscript is organized as follows. In
Sec.~\ref{tools}, we start our study by introducing both quantum
discord and its global version. In Sec.~\ref{results} we then move
to the description of a set of physically relevant interacting
quantum many-body models that will be studied against the content
of global quantum correlations of equilibrium states at
temperature $T\neq{0}$ and present our results. Finally, in
Sec.~\ref{conclusions} we draw up our conclusions and discuss a
few open questions that are left to a future addressing.

\section{Tools for quantifying quantum correlations}
\label{tools}

In this Section we introduce the fundamental mathematical tools
used in our study. We recall the definition of global discord
given in Ref.~\cite{RulliArX11}, and present a more agile
expression for the case of multipartite qubit systems. For the
sake of completeness we briefly review the original formulation of
quantum discord valid in the bipartite scenario.

\subsection{Quantum Discord}
\label{sec:qd}

We begin by reminding that, as originally proposed
in~\cite{Ollivier}, quantum discord is linked to the discrepancy
between two quantum extensions of the concept of conditional
entropy that are classically equivalent. Let us
consider a bipartite system described by the density operator
$\rho_{AB}$ with $\rho_{A}$ ($\rho_{B}$) denoting the reduced
state of system A (B). The total correlations between A and B are
quantified by the mutual information
\begin{equation}
I(\rho_{AB})=S(\rho_{A})-S(\rho_{A}|\rho_{B}),
\end{equation}
where $S(\rho_{A})=-\mathrm{Tr}[\rho_{A}\log_2\rho_{A}]$ is the
von Neumann entropy and
$S(\rho_{A}|\rho_{B})=S(\rho_{AB})-S(\rho_B)$ is the conditional
entropy. By using a measurement-based approach, a second
definition of conditional entropy can be formulated. The
application of a local projective measurement,
described by the set of projectors $\{\hat{\Pi}_B^j\}$ on part B
of the system, results in the conditional post-measurement density
operator
$\rho_{AB|j}=(\openone_A\otimes\hat{\Pi}_B^j)\rho_{AB}(\openone_A\otimes\hat{\Pi}_B^j)/p_j$,
where
$p_j=\mathrm{Tr}[(\openone_A\otimes\hat{\Pi}_B^j)\rho_{AB}]$ is
the probability associated with the measurement outcome $j$. We
can thus define the measurement-based conditional entropy
$S(\rho_{AB}|\hat{\Pi}_B^j)=\sum_j p_j S(\rho_{A|j})$ with
$\rho_{A|j}=\mathrm{Tr}[\hat{\Pi}^j_B\rho_{AB}]/p_j$, which leads
us to the so-called one-way classical information~\cite{Henderson}
\begin{equation}
J(\rho_{AB})=S(\rho_A)-S(\rho_{AB}|\hat{\Pi}_B^j).
\end{equation}
The difference between quantum mutual information and classical
correlations, minimized over the whole set of orthogonal
projective measurements performed on B, defines quantum discord as
\begin{equation}
\label{discord}
{\cal D}^{B\rightarrow A}(\rho_{AB})=\inf_{\{\hat{\Pi}_B^j\}}[I(\rho_{AB})-J(\rho_{AB})].
\end{equation}

By noticing that the original definition of
discord~\cite{Ollivier} can be rewritten in terms of relative
entropy
$S(\rho_1||\rho_2)=\mathrm{Tr}[\rho_1\log_2\rho_1]-\mathrm{Tr}[\rho_1\log_2\rho_2]$
between two generic states $\rho_1$ and
$\rho_2$~\cite{RulliArX11} and by symmetrizing its definition through the introduction of bilateral measurements $\hat\Pi^j_A\otimes\hat\Pi^k_B$~\cite{io}, we introduce 
\begin{equation}\label{Dsym}
\mathcal{D}^{AB}(\rho_{AB})=\min_{\{\hat \Pi^j_A\otimes\hat \Pi^k_B\}}[S(\rho_{AB}||\Pi(\rho_{AB})]
-\sum_{j=A,B}S(\rho_{j}||\hat{\Pi}_j(\rho_{j}))
\end{equation}
with $\hat\Pi(\rho_{AB})=\sum_{j,k}(\hat\Pi^j_A\otimes\hat\Pi^k_B)\rho_{AB}(\hat\Pi^j_A\otimes\hat\Pi^k_B)$
Eq.~(\ref{Dsym}) expresses discord as the difference between the content of quantum correlations ascribed to a multi-local measurement process and the sum of the 
relative entropies for each reduced state of the system. The minimisation is required, clearly, to remove any dependence on the local measurement bases. The absence of global quantum correlations would make Eq.~\eqref{Dsym} identically null.
\subsection{Global Quantum Discord} 

Eq.~(\ref{Dsym}) is the starting point for the formulation of global discord (GD)~\cite{RulliArX11}
\begin{equation}
\label{GQD}
\mathcal{GD}(\rho_{T})=\min_{\{\hat\Pi^k\}}\left\{S\left(\rho_{T}||\hat\Pi(\rho_{T})\right)-\sum_{j=1}^N S\left(\rho_{j}||\hat\Pi_j(\rho_{j})\right)\right\},
\end{equation}
which quantifies the global content of non-classical correlations in the state $\rho_T$ of an $N$-party system. Here 
$\rho_j=\mathrm{Tr}'\left[\rho_T\right]$ is the reduced state of qubit $j$ [we use $\textrm{Tr}'$ for the trace over all the qubits but the $j^{\textrm{th}}$], $\hat\Pi_j(\rho_{j})=\sum_{l}\hat\Pi_{j}^{l}\rho_{j}\hat\Pi_{j}^{l}$, $\hat\Pi(\rho_{T})=\sum_k \hat\Pi^k \rho_{T} \hat\Pi^k$, $\hat \Pi^k=\otimes^N_{l=1}\hat \Pi^{k_l}_{l}$, and $k$ stands for the string of indices $(k_1\dots k_N)$. The minimization inherent in Eq.~(\ref{GQD})  is performed over all possible multi-local projectors $\hat \Pi^k$.
In Ref.~\cite{RulliArX11} it is shown that ${\cal GD}(\rho_{T})\ge0$, its maximum value depending on the dimension of the total Hilbert space at hand. Recently, a monogamy relation relating global quantum discord in a multipartite setting and pairwise correlations evaluated by quantum discord has been introduced in Ref.~\cite{Braga}.

The explicit computation of the formula in Eq.~(\ref{GQD}) is in general a difficult problem. However, the task can be greatly simplified by writing the multi-qubit projective operators as $\hat{\Pi}^k=\hat{{\mathcal R}}\ket{{\bf k}}\bra{{\bf k}}\hat{{\mathcal R}}^{\dag}$. Here $\left\{\ket{{\bf k}}\right\}$ are separable eigenstates of $\hat\Sigma_z=\otimes^N_{j=1}\hat\sigma_j^z$ with $\hat{\sigma}_j^q$ the $q=x,y,z$ Pauli operator, and $\hat{{\mathcal R}}$ is a local multi-qubit rotation $\hat{{\mathcal R}}=\otimes^N_{j=1}\hat{R}_j(\theta_j,\phi_j)$ with $\hat{R}_j(\theta_j,\phi_j)=\cos\theta_j\hat\openone+i\sin\theta_j\cos\phi_j\hat\sigma_y+i\sin\theta_j\sin\phi_j\hat\sigma_x$ the rotation operator (of angles $\theta$ and $\phi_j$) acting on the $j$-th qubit. Analogously, the set of local projective operators on the $j$-th qubit is written as $\hat{\Pi}_j^l=\hat{R}_j\ket{l}\bra{l} \hat{R}_j^{\dag}$ with $\ket{l}$ ($l=0,1$) the eigenstates of $\hat{\sigma}_j^z$ and where, for convenience, we have dropped the dependence of the rotation operators on their respective angles.
As shown in some details in the Appendix, the introduction of these quantities allows us to reformulate GD as
\begin{equation}\label{GDsimp}
\begin{aligned}
\mathcal{GD}(\rho_{T})&=\min_{\{\hat\Pi^k\}}\left\{\sum_{j=1}^{N}\sum_{l=0}^1\tilde{\rho}_j^{ll}\log_2\tilde{\rho}_j^{ll}-\sum_{k=0}^{2^N-1}\tilde{\rho}_T^{kk}\log_2\tilde{\rho}_T^{kk}\right\}\\
&+\sum_{j=1}^N S(\rho_j)-S(\rho_T)
\end{aligned}
\end{equation}
with $\tilde{\rho}_T^{kk}=\bra{{\bf k}}\hat{\mathcal{R}}^{\dag}\rho_T\hat{\mathcal{R}}\ket{{\bf k}}$ and $\tilde{\rho}_j^{ll}=\bra{l}\hat{R}_j^{\dag}\rho_j \hat{R}_j\ket{l}$. Despite its involved form, Eq.~(\ref{GDsimp}) greatly reduces the computational efforts needed to evaluate ${\cal GD}(\rho_{T})$.

\section{Quantum correlations and criticality in quantum spin chains}
\label{results} To examine the qualitative and quantitative
features of quantum correlations in spin systems, we will focus on
three different models: the {\it Ising}, {\it cluster-Ising} and
the {\it XX} models. Here we are interested not only in the zero
temperature case, but also in exploring the thermal effects on
such finite sized quantum systems that exhibit critical behavior.
For the size of the systems we will be considering, the explicit
thermal state can be directly calculated via its canonical
ensemble and is given by the Gibbs state [throughout this
manuscript we take units such that $\hbar=k_B=1$]:
\begin{equation}
\label{gibbs}
\varrho(T)=\frac{e^{-\hat{\mathcal{H}}/{T}}}{\mathcal{Z}}
\end{equation}
with $\hat{\mathcal{H}}$ the Hamiltonian describing the
interaction, $T$ the effective temperature, and
$\mathcal{Z}=\mathrm{Tr}[e^{-{\hat{\cal H}}/{T}}]$ the partition
function.

\subsection{Transverse field Ising Model}
\label{ising} We start our analysis considering the quantum Ising
model in the zero-temperature case. The behavior of bipartite and
global correlations in the transverse spin-$1/2$ Ising model has
attracted considerable interest so far.
Entanglement~\cite{osborne,buzek,igloi1}, non-locality~\cite{campbell},
and bipartite quantum discord~\cite{campbell2} have been studied
for this model. More recently, the scaling of entanglement
spectrum of a finite-size spin-$1/2$ Ising chain near its critical
point has been studied~\cite{gabriele}.

Here, in line with some of the studies mentioned above, we shall
consider a one dimensional system with periodic boundary
conditions. The Hamiltonian for a chain of $L$ spin-$1/2$
particles reads
\begin{equation}
\hat{\mathcal{H}}_I=-J \sum_{i=1}^L \hat\sigma_i^x\hat\sigma_{i+1}^x + B\sum_{i=1}^L \hat\sigma_i^z
\label{isingmodel}
\end{equation}
with the condition $L+1\equiv1$. 
In the limit $B/J\rightarrow0$, the ground state of this model is locally equivalent to an $L$-spin GHZ state~\cite{buzek}. 
As $B$ increases, the entanglement in the ground state soon disappears, as the spins tend to align along the direction set by the magnetic field and, in the thermodynamic limit, the system undergoes a quantum phase transition at $B/J=1$. The nonlocal nature of the quantum correlations within the ground state of this model has been studied in~\cite{campbell} and found to be extremely sensitive to temperature, a feature shared with the entanglement. Needless to say, this does not imply that all non-classical features in the correlations shared among all the spins disappear with temperature and we shall demonstrate that global quantum discord is indeed able to signal the structural changes in the sharing of quantum correlations even at $T\neq0$. 

Technically, the evaluation of GD for the Ising model enjoying the
symmetries mentioned above offers room for a few interesting
considerations. First, as the model in Eq.~\eqref{isingmodel} is
characterized by a real symmetric Hamiltonian, GD is completely
independent on the set of $\phi_j$ angles, which do not play any
role in the minimization necessary to calculate Eq.~\eqref{GQD}.
Second, as we have taken periodic boundary conditions, the system
is translational invariant. Such invariance has consequences on
the relation among $\theta_j$, making GD invariant under cyclic
permutations of such variables. Finally, we have gathered
numerical evidence of even a higher degree of symmetry of the GD
function, at all  lengths and values of the temperature, in that
the optimal $\theta_j$ all take the same value
$\overline{\theta}$, which depends on the magnetic field. We have $\overline{\theta}=0$ [$\pi/4$]  for small [large] values of
$B$. The transition between these two values is sharp and happens
in proximity of the critical point, thus showing how the changes
induced at criticality are reflected in the structure of GD. We
have run our simulations by minimizing over all the possible
different angle configurations, finding perfect agreement with the
results corresponding to this explicit choice of angles.

\begin{figure}[t]
\centering
\includegraphics[width=0.99\linewidth]{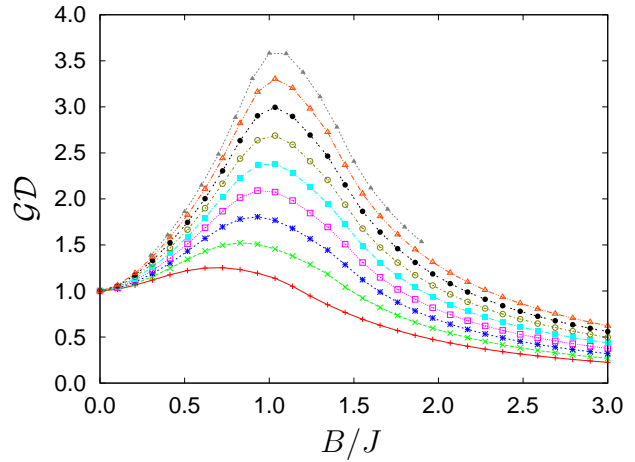}
\caption{$\cal{GD}$ for the Ising model at zero temperature. From bottom to top curve, $L$ goes from 3 to 11 spins. At $B=0$ the ground state of the spin model is a GHZ (corresponding to the ferromagnetic case with no symmetry breaking), thus giving ${\cal GD}=1$~\cite{RulliArX11}. In the paramagnetic configuration ($B\gg{J}$), ${\cal GD}$ goes to zero together with any non trivial spin correlation.}
\label{IsingT0}
\end{figure}

Fig.~\ref{IsingT0} shows the amount of quantum correlations, quantified by GD, as a function of the ratio $B/J$ between the magnetic field intensity and the Ising interaction constant. We study rings with $L=3,\dots,11$ whose GD curves share the same value at zero magnetic field. This agrees with the ground state being an $L$-spin GHZ,  for which ${\cal GD}=1$ regardless of the system size. As $B/J$ tends to 1 the global discord increases reaching a maximum at different positions depending on the length of the chain. In the paramagnetic phase achieved for $B\gg{J}$, all the spins align along the direction of the magnetic field, so that all quantum correlations disappear. At smaller values of $B$, however, the global sharing of quantum correlations undergoes significant changes which result in the appearance of a maximum, whose height and position is a clear function of the size of the system.

\begin{figure}[b!]
\centering
{\bf (a)}\\
\includegraphics[width=0.8\linewidth]{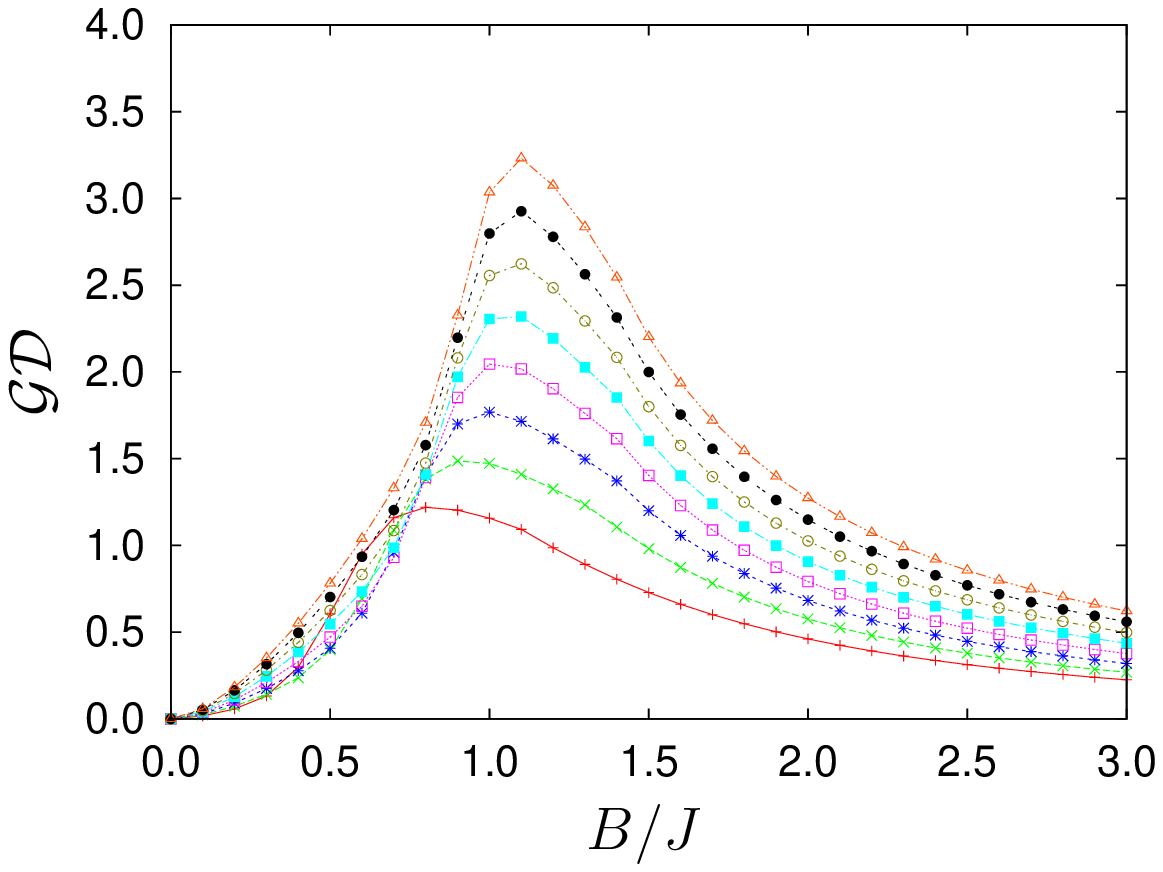}\\
{\bf (b)}\\
\includegraphics[width=0.8\linewidth]{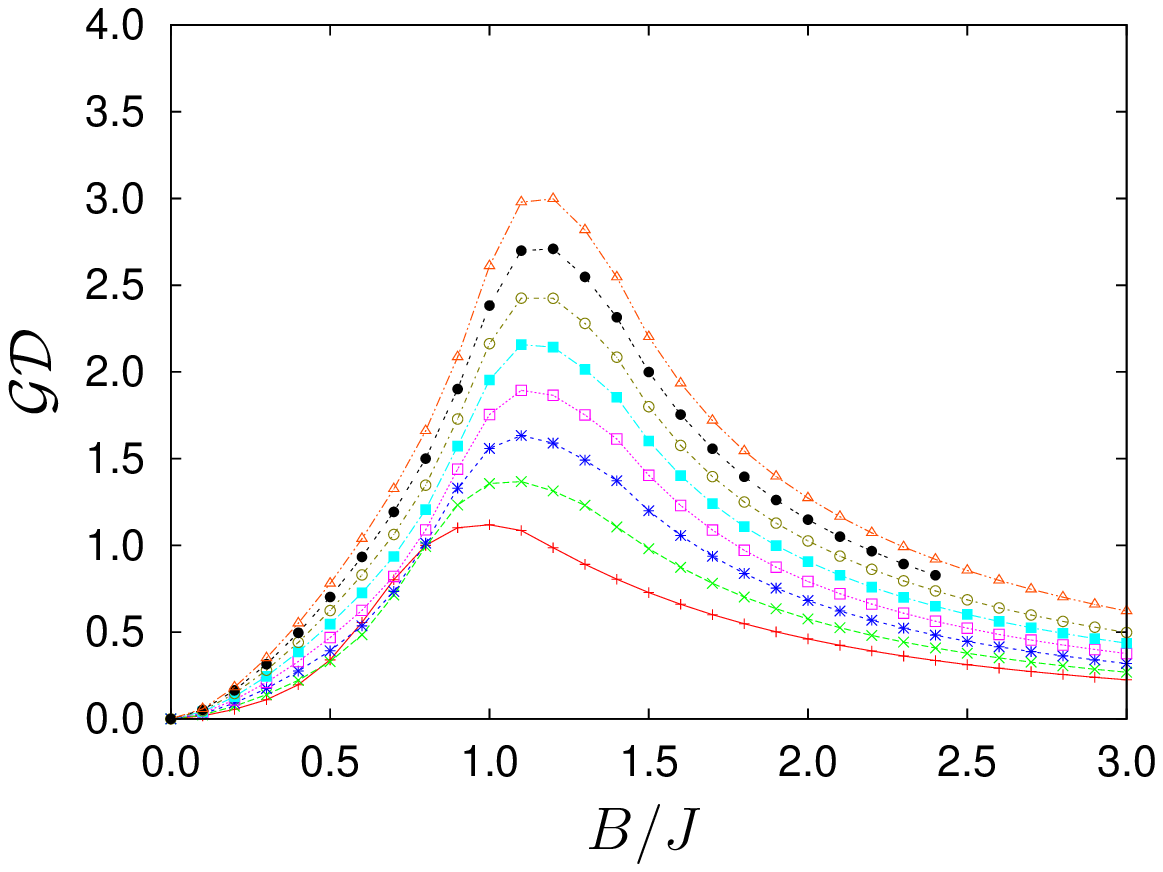}
\caption{${\cal GD}$ for Ising rings containing from 3 to 10 spins at non-zero temperature: ({\bf a}) $T=0.05$, ({\bf b}) $T=0.1$. GD is null at zero magnetic field. For increasing temperatures the maximum values of the curves decrease.}
\label{IsingT}
\end{figure}

The core part of our analysis consists of the study of the changes in the behaviour of GD for rings prepared in thermal states. We consider two different cases with effective temperature 
equal to $T=0.05$ [cf. Fig.~\ref{IsingT}~({\bf a})] and $T=0.1$ [cf. Fig.~\ref{IsingT}~({\bf b})]. At non-zero temperature, the quantum correlations that are present in the ground state at $B=0$ are destroyed and ${\cal GD}(\rho_T)=0~\forall{L}$. This is due to the fact that as $B\to0$ the ground and first excited states approach degeneracy. These states are both of GHZ-type~\cite{buzek} and therefore even small $T$ is sufficient to completely mix the ground and first excited states and destroy all quantum correlations~\cite{campbell}. Overall the height of the curves decreases with increasing temperature. At low temperatures [cf. Fig.~\ref{IsingT}~({\bf a})], the position of the maxima of ${\cal GD}(\rho_T)$ is extremely close to those at $T=0$, while higher temperatures induce a shift in the maxima of each curve
~ [the effect is already visible in Fig.~\ref{IsingT}~({\bf b})].
The optimal angles for the GD of thermal states are the same as those for $T=0$.

\begin{figure}[t]
\centering
\includegraphics[width=0.99\linewidth]{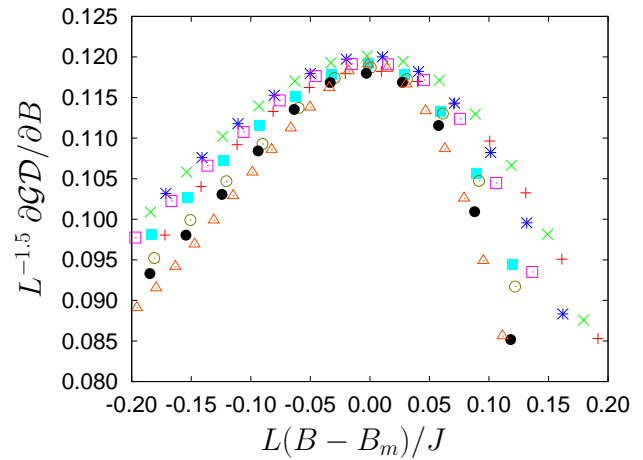}
\caption{Derivative of GD with respect to the magnetic field for the Ising model at zero temperature. The data points (for $L=4,\dots,11$) are scaled according to the number of spins of the rings. $B_m$ is the critical value of the magnetic field for a finite-size chain of length $L$. Close to the critical point the curves at all values of $L$ collapse to a common function, witnessing universality.}
\label{DerivIsing}
\end{figure}

In line with the studies performed on the scaling of entanglement in quantum spin systems~\cite{latorre}, it is interesting to study the way global quantum correlations scale against the number of elements of the multipartite systems that we are addressing here. To this aim, we perform a finite-size scaling analysis of GD in the proximity of the critical point $B_m/J$ for a finite-size transverse Ising model at $T=0$. 
We thus study the derivative of ${\cal GD}$ with respect to $B$ finding that, in proximity of the finite-size critical point, it is a function of $L^{\nu}(B-B_m)/J$ satisfying the scaling ansatz~\cite{Fisher72,tomasello}:
\begin{equation}
\frac{\partial}{\partial B}{\cal GD}=L^{-\omega}f[L^{\nu}(B-B_m)/J]
\end{equation}
where $\nu=1$ is the correlation length divergence critical exponent of the corresponding Ising universality class. There is clear indication of data collapse already for very small lengths ($L\le 11$) and we obtain $\omega\approx-1.5$ while $f(x)$ is approximately quadratic for $x\sim0$  [cf. Fig.~\ref{DerivIsing}]. This is quite a remarkable result, as it shows that GD appears to scale with universal critical exponents in the proximity of a quantum phase transition.

\subsection{Cluster-Ising model}
\label{cluster}

We now consider a model, recently proposed and studied in Ref.~\cite{Son}, which combines competing effects coming from an antiferromagnetic Ising and a three-body cluster interaction according to the Hamiltonian (with periodic boundary conditions)
\begin{equation}\label{3bodymodel}
\hat{\mathcal{H}}_{CI}=-J\sum^L_{i=1}\hat\sigma_{i-1}^x \hat\sigma_{i}^z\hat\sigma_{i+1}^x + \lambda \sum^L_{i=1} \hat\sigma_i^y \hat\sigma_{i+1}^y.
\end{equation}
The three-body term in Eq.~(\ref{3bodymodel}) is responsible for the setting of long-range entanglement, which has been recently related to topologically ordered states~\cite{Chen}, while the second term tends to localise entanglement to nearest-neighbor pairs of spins. Such competition makes the model undergo a second-order quantum phase transition at $J/\lambda=1$ with the ground state of the system passing from an Ising antiferromagnetic phase (for $J/\lambda\ll{1}$) to a {\it cluster-like} one (achieved at $J/\lambda\gg1$) endowed, as said above, with long-range entanglement and topological order. Such transition, which is not in the Ising universality class, has been characterized in Ref.~\cite{Son,Smacchia,clusterref} by means of a global geometric measure of entanglement and entanglement spectrum.

An interesting point to notice is that, for a cluster-Ising model
at non-zero temperature, neither the two-spin nor the multipartite
entanglement (as measured by tangle) is able to signal the quantum
phase transition, as they are either identically null (two-spin
entanglement) or equal to a constant (multipartite tangle). Here
we will make use of global quantum discord to study the occurrence
of critical structural changes in the correlation-sharing
structure of the model
even at finite temperature, showing the effectiveness of GD in the task of
revealing such modifications at criticality. 
Technically, the problem of finding the minimum in Eq.~\eqref{GQD}
is more difficult to tackle than in the Ising case. In fact, the
model is characterized by a lower degree of symmetry (due to the
presence of all the Pauli spin operators in
Eq.~\eqref{3bodymodel}) which forces us to minimize
Eq.~\eqref{GDsimp} using all angles  $\{\theta_j\}$ and
$\{\phi_j\}$. For $J/\lambda\ll 1$ (Ising phase) the optimal
angles are $\{\theta_j=\pi/4\}$ and $\{\phi_j=\pi/2\}$ analogously
to the ferromagnetic phase of the Ising model (the value of the
phases $\phi_j$ is due to the different Ising coupling of this
model). In the opposite regime $J/\lambda\gg 1$ (cluster phase)
the optimising angles depend on the number of spins.

\begin{figure}[t]
\centering
\includegraphics[width=0.9\linewidth]{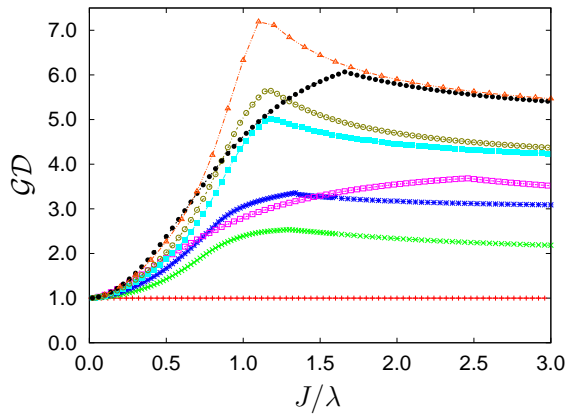}
\caption{${\cal GD}$ for the cluster-Ising model at zero temperature. We took $L=3,\dots,10$, growing from the bottom to the top curve as they appear in the $J/\lambda\gg{1}$ part of the plot. In the second case ${\cal GD}$ for cluster states with an increasing number of spins starting from 3 is equal to 1, 2, 3, 3, 4, 4.}
\label{ClusterT0}
\end{figure}

We start the description of our result by analyzing
Fig.~\ref{ClusterT0}, where we plot the GD as a function of
the ratio $J/\lambda$ for $L=3,\dots,10$ and at $T=0$. In the
limit of vanishing cluster-like contribution, the model contains
the Ising two-body interaction and we correspondingly recover the
results obtained in Sec.~\ref{ising} for zero transverse magnetic
field: the ground state is locally equivalent to an $L$-spin GHZ
state and ${\cal GD}=1$ regardless of the size of the chain. In
the opposite asymptotic regime, where the three-body model
dominates over the Ising term, the ground states are cluster
states on an $L$-site ring lattice~\cite{clusterreview}. Even at
moderately large choices of $J/\lambda$, the value taken by GD
agrees very well with the expectations for size-$L$ cluster
states, which are equal to $1,2,3,3,4,4$ for $L$ going from 3 to
8, respectively.
Away from such limits, GD behaves in a peculiar way with the size
of the system. Most of the cases that we have considered in our
analysis display a non-monotonic behavior with a peak occurring in
proximity of the critical point. However, among the values of $L$
considered in our calculations, the cases of $L=3$ and $L=6,9$
behave differently, with the GD being practically constant or with
a sharper maximum at values of $J/\lambda$ significantly away from
the critical point. Moreover, these special cases give rise to a
few crossings with the curves associated with both lower and
larger rings (for instance, the curve corresponding to $L=6$
crosses both those for $5$ spins and $7$ spins).

We believe that the occurrence of such {\it pathological} behavior
when $L$ is a multiple of $3$ should not be regarded as
accidental,  but rather as a signature of the distinctive features
of the cluster-Ising model for these lengths of the system. In
fact, the study (at finite size) conducted in Ref.~\cite{Smacchia}
shows that, differently from the thermodynamic limit, the $x$- and
$z$-correlations in the model, as well as the magnetization along
$z$ axis, vanish for $L$ that is a multiple of 3
(Ref.~\cite{Smacchia} discusses explicitly the case of $L= 6$ and
$12$). Although, given the complexity of the task, it is
implausible to formulate an analytic expression of GD from which
the role played by such correlations can be clearly extracted, we
conjecture their relevance in the determination of the functional
form of GD at moderate values of $J/\lambda$, where the
differences with respect to any other system-size appear to be
more striking. In turn, given the agreement between the calculated
GD and the predictions valid for the ground state of the
Ising and cluster models at all values of $L$, this analysis
suggests that the above mentioned correlators are not heavily
relevant for the calculation of GD deep in both such phases. While
this observation  could help in studying this figure of merit
analytically, a less phenomenological approach to this interesting
points goes beyond the scopes of this work and remains to be
addressed in further studies on this matter.

\begin{figure}[t]
\centering
{\bf (a)}\\
\includegraphics[width=0.8\linewidth]{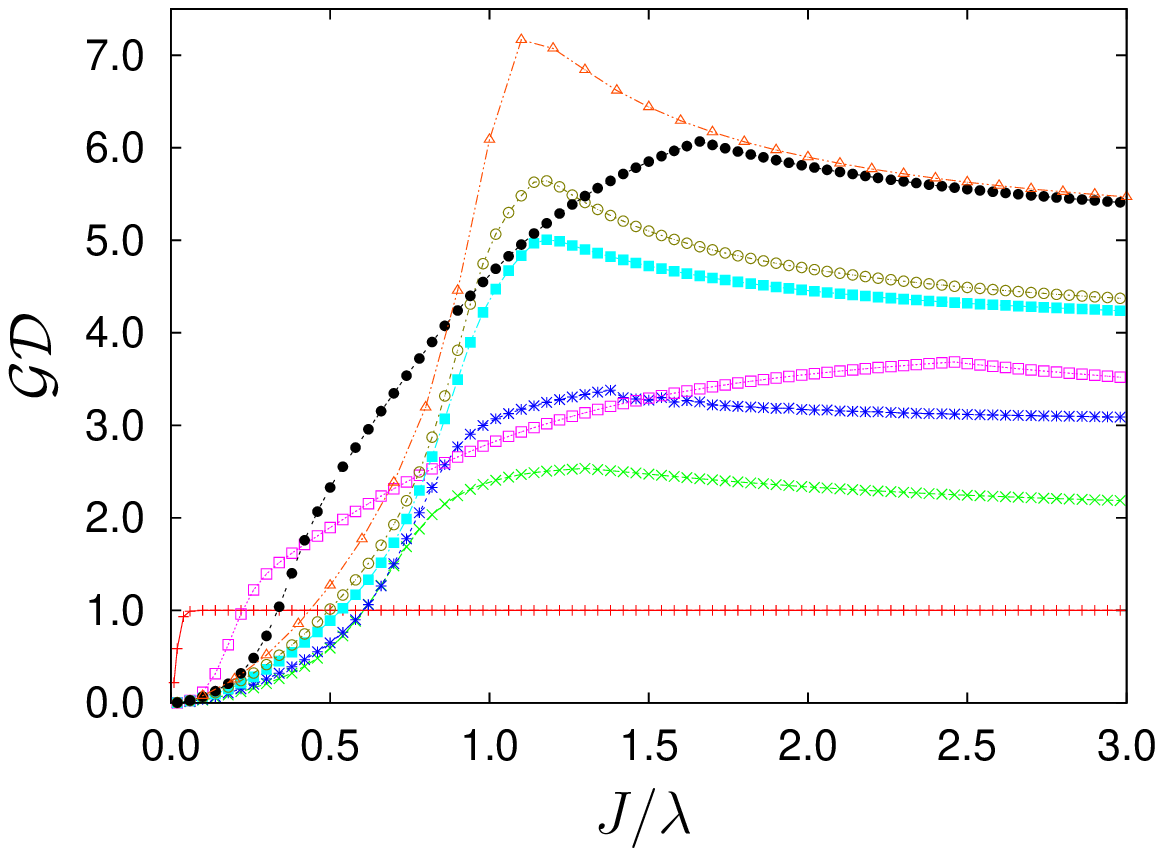}\\
{\bf (b)}\\
\includegraphics[width=0.8\linewidth]{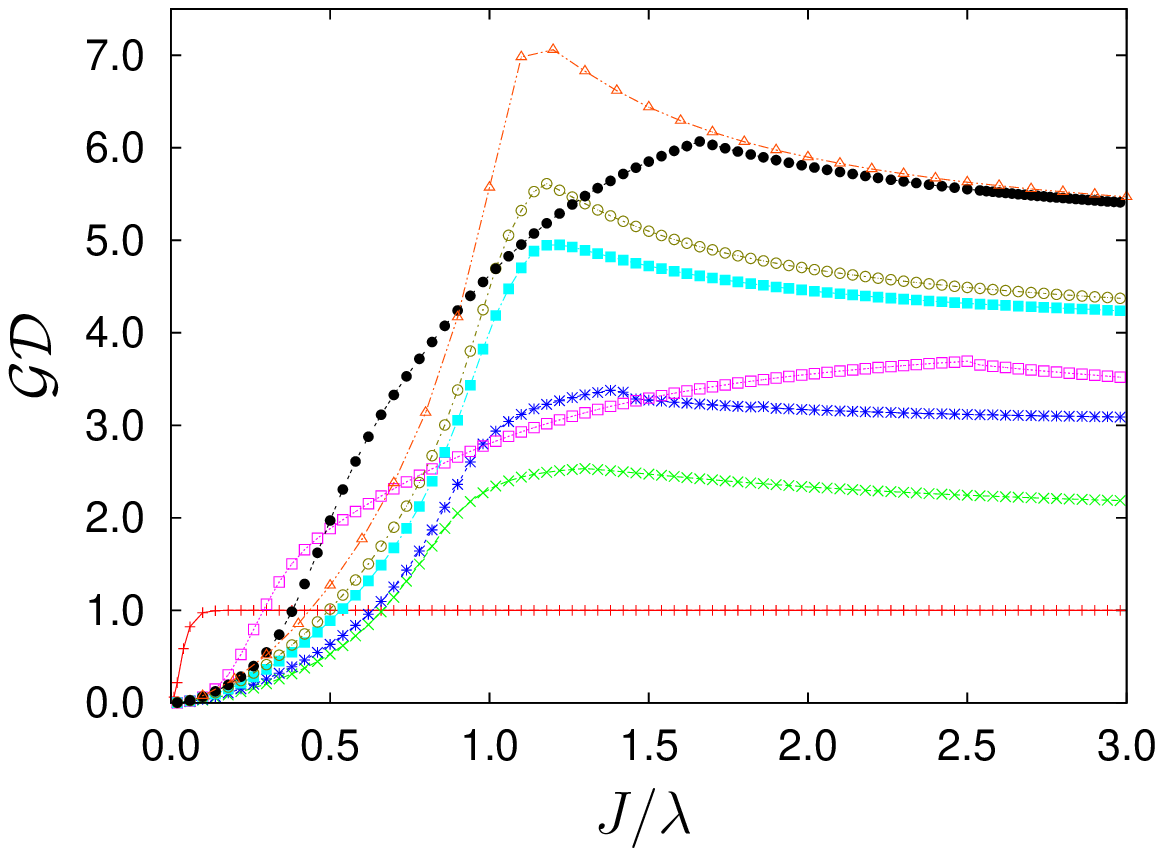}
\caption{${\cal GD}$ for the cluster-Ising model at non-zero temperature. We took $L=3,\dots,10$, growing from the bottom to the top curve as they appear in the $J/\lambda\gg{1}$ part of the plot. In panel ({\bf a})  [{\bf (b)}] we took $T=0.05$ [$T=0.1$].}
\label{ClusterT}
\end{figure}

We conclude our analysis of this model by addressing now the case
of $T\neq0$, as done in Fig.~\ref{ClusterT}. While GD vanishes in the antiferromagnetic phase, it persists to the
effects of temperature in the cluster one. This is in line with
what has been found for thermal cluster states, for which an
$L$-dependent critical temperature exists, below [above] which
distillable and long-range [bound] entanglement is found in the
cluster state~\cite{thermalcluster,commentocluster}.

Even more strikingly, though, the structure observed at $T=0$
survives, qualitatively unaltered, at $T\neq0$. Actually, the
quantitative differences between the results associated with the
$L$-spin ground state and the corresponding thermal equilibrium
state are negligible [in terms of both the position of the maximum
of GD on the $J/\lambda$ axis and their actual value, cf.
Fig.~\ref{ClusterT} {\bf (a)} and {\bf (b)}] even at values of $T$
for which the GD of an Ising chain was found, in Sec.~\ref{ising},
to be sensibly different from that of the $T=0$ case. Needless to say,
at much larger values of the temperature the peaks close to the
critical point are smeared out into a broad and flat GD curve.
This demonstrates the claimed effectiveness of the figure of merit
addressed here in signalling the effects of criticality on the
sharing of quantum correlations in thermal-equilibrium states,
thus reinforcing the significance of the analysis conducted so far
along the lines of combining quantum many-body physics and
discord-related quantifiers~\cite{varie,tomasello,campbell2}.

\subsection{Open-chain XX model}
\label{xx}

\begin{figure}[t]
\centering
\includegraphics[width=0.9\linewidth]{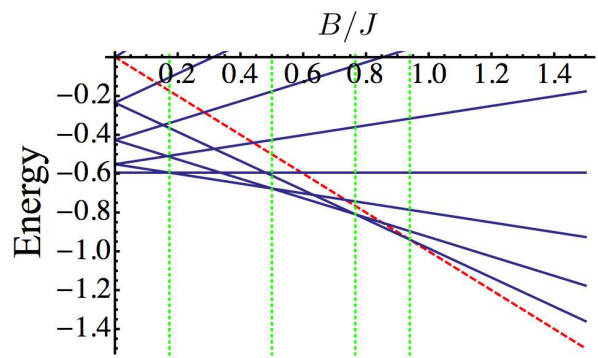}
\caption{Spectrum of the model defined in Eq.~\eqref{XXmodel} against the ratio $B/J$ for $L=8$ spins. The red dashed line represents the final ground state of the model after all energy crossings. Dashed vertical lines identifies the points of occurrence of level crossings in the model.}
\label{spettro}
\end{figure}

Finally, we address the case  of an open-ended chain interacting via an XX term in the presence of a transverse magnetic field
\begin{equation}
\hat{\mathcal{H}}_{XX}=-\frac{J}{2}\sum^{L-1}_{i=1} (\hat\sigma_i^x \hat\sigma_{i+1}^x +\hat\sigma_i^y \hat\sigma_{i+1}^y) - B \sum^{L}_{i=1} \hat\sigma_i^z.
\label{XXmodel}
\end{equation}
As studied in Ref.~\cite{sonamico} and shown for $L=8$ in
Fig.~\ref{spettro}, the energy spectrum of this model is quite
rich and encompasses quite an interesting series of crossings
among its energy eigenstates occurring as the ratio $B/J$ is
varied. Correspondingly, the ground state of the system changes
and can be classified in terms of the number of such level
crossings. In the thermodynamic limit $L\to\infty$, $\hat{\cal
H}_{XX}$ exhibits a  Berezinskii-Kosterlitz-Thouless (BKT)
transition at $B/J=1$ from a critical phase, characterized by
in-plane quasi-long range order, to a paramagnetic one.
Correspondingly, two-spin quantum entanglement (as measured by
concurrence) reduces as the distance between the spins is
increased, and vanishes at the critical point, leaving a  fully
factorized ground state.

\begin{figure}[b]
\centering
\includegraphics[width=0.9\linewidth]{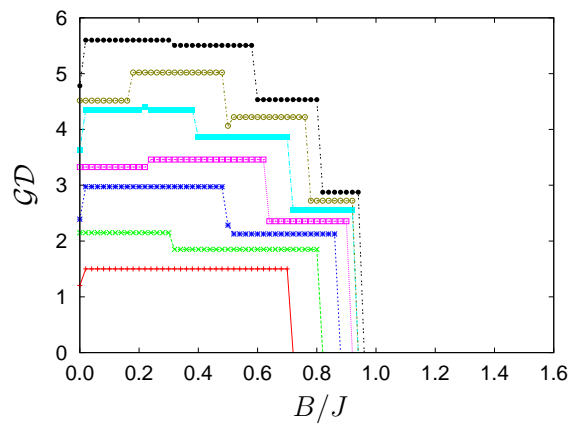}
\caption{${\cal GD}$ for the XX model at  $T=0$. We have taken $L=3,\dots,9$ (from bottom to top curve).}
\label{XXT0}
\end{figure}

\begin{figure}[!t]
\centering{\bf (a)}\\
\includegraphics[width=0.9\linewidth]{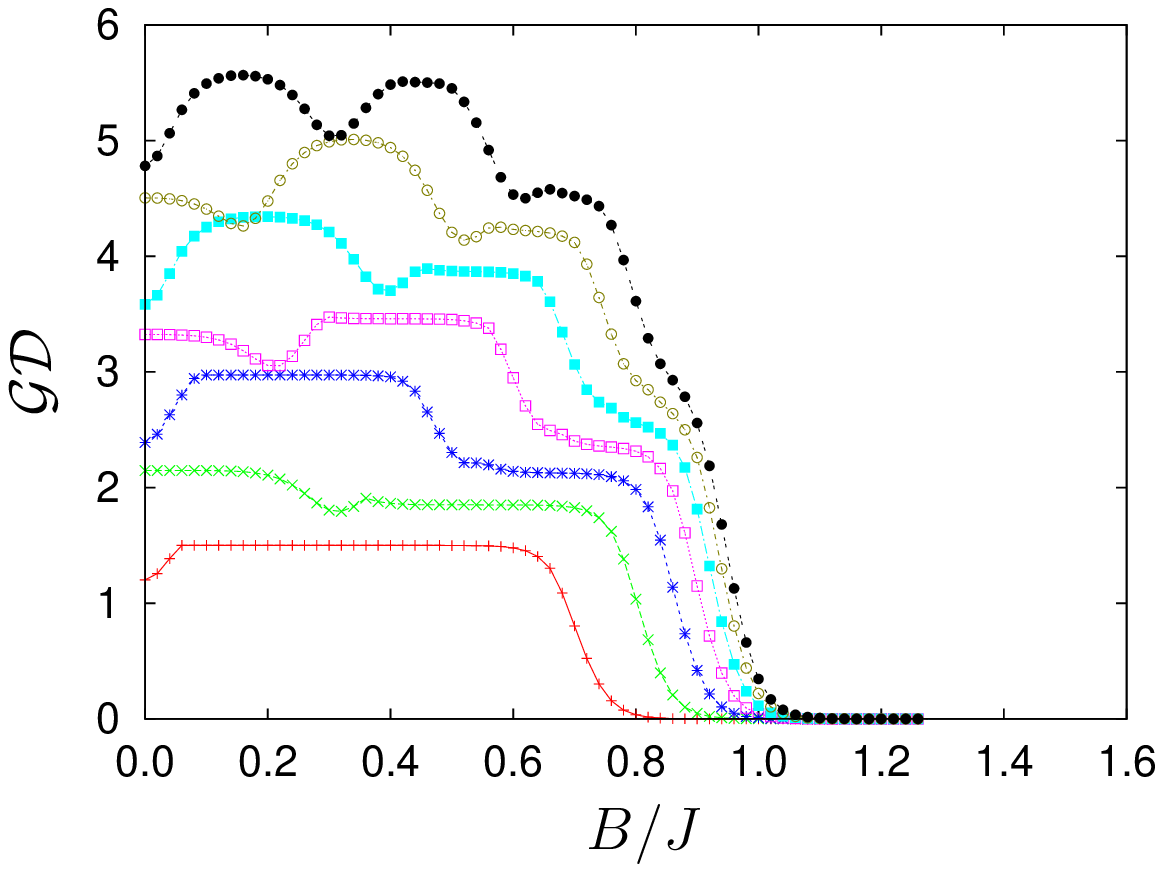}\\
\centering{\bf (b)}\\
\includegraphics[width=0.9\linewidth]{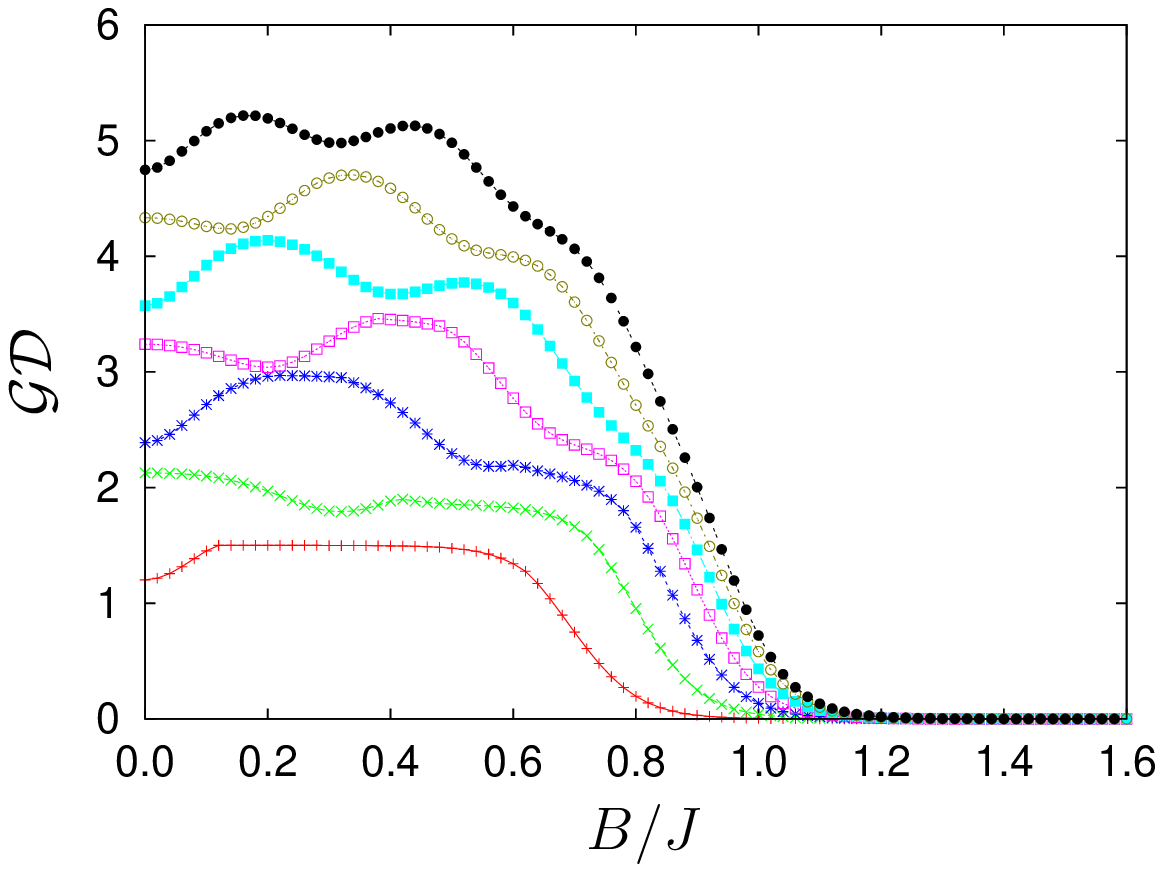}
\caption{${\cal GD}$ for the XX model at non-zero temperature. In panel ({\bf a}) [{\bf (b)}] we have taken $T=0.05$ [$T=0.1$]. In both panels we have $L=3,\dots9$ (from bottom to top curve, respectively).}
\label{XXT}
\end{figure}

Fig.~\ref{XXT0} shows GD as a function of the
external magnetic field over the two-body interaction constant. As
discussed above, for high magnetic field the system is in the
paramagnetic phase and no correlations are present. For low
magnetic field the GD displays a step-wise behavior, jumps
occurring in correspondence of the level-crossings that redefine
the ground state of the system (which are evident from the
spectrum of the model). That is, GD tracks the structural
changes in the ground state of the spin system as $B/J$ varies.
Moreover, as $L$ grows, GD goes to zero at
values of $B/J$ increasingly closer to the critical point. Already
for the modest value of $L=9$, the difference between the value of
$B/J$ at which ${\cal GD}=0$ and $B/J=1$ is only $5\%$.
A non-zero temperature smoothens the sharpness of the jumps
occurring in GD and reduces its amplitude as shown in
Fig.~\ref{XXT}. Yet, the series of level crossings at which the
ground state changes, as well as the BKT point are still clearly
visible as dips between quasi-plateaux and a gradual yet  quick
decrease of ${\cal GD}$, respectively, thus proving the
effectiveness of GD as a figure of merit for
signalling criticality at $T\neq0$. It should be noted that, for $B=T=0$, the ground state is doubly degenerate. In order to evaluate 
GD we have thus taken the linear combination of such degenerate states that smoothly provides the values associated to $T\neq0$,
where the degeneracy is lifted. 

\section{Conclusions}
\label{conclusions}

We investigated the behavior of a measure of global quantum
correlations in three finite-size one-dimensional quantum
many-body systems close to their critical points. We have
manipulated the expression for global quantum discord so as to
adapt it to the case of quantum spin chains. This has allowed us
to study the global quantum discord in thermal states of
moderately large quantum spin models, demonstrating its
effectiveness in spotting the critical changes in the state of the
system as a function of relevant parameters. Furthermore, for the
Ising model we have been able  to put forward evidences of a
finite-size scaling behavior characterized by universal critical
exponents of the Ising universality class. For the cluster-Ising
model, GD provides an alternative and powerful tool for the
signalling of criticality superior to entanglement measures in
both reduced and global forms. Finally, for an open XX chain, we
have been able to track the discrete number of structural changes
for the ground state as the transverse magnetic field varies.

The relation between the symmetries in a given model and the calculation of the correlations present is highly non-trivial~\cite{amicoopt}. Our analysis sheds light also on technical aspects related to the
calculation of global discord in multipartite spin systems enjoying
some degree of symmetry.
For the transverse Ising model with periodic boundary conditions, we have
provided strong numerical evidence that identical local projections should
be implemented in order to attain the global maximum inherent in the definition of GD.
Moreover, the azimuthal angles $\phi_j$ are shown to be immaterial for this task. 
Differently, for the cluster-Ising model in
Eq.~(\ref{3bodymodel}), deep in the cluster phase, more
complicated combinations of minimizing angles are found. Although,
it would be interesting to find a relation between the symmetries
of the model in consideration and the angles minimizing the global
discord, a comprehensive solution seems highly non-trivial and
goes beyond the scope of this study.

\section*{Acknowledgments}
We are grateful to L. Amico, G. L. Giorgi, V. Vedral, and A. Xuereb for invaluable discussions.
We acknowledge funding from the EU under the Marie Curie IEF Fellowship scheme, the European Commission, the
European Social Fund, the Region Calabria through the program POR
Calabria FSE 2007-2013 - Asse IV Capitale Umano-Obiettivo Operativo
M2, Science Foundation Ireland under project number
10/IN.1/I2979, the UK EPSRC through a Career Acceleration Fellowship and  the ``New Directions for EPSRC Research Leaders" initiative (EP/G004579/1).

\section*{Appendix}

Here we present the main steps of the derivation of the expression
of the global discord for N spin-system presented in
Eq.~(\ref{GDsimp}). The starting point is writing the multi-local
projectors as $\hat{\Pi}^k=\hat{{\mathcal R}}\ket{{\bf
k}}\bra{{\bf k}}\hat{{\mathcal R}}^{\dag}$ where we remind
that $\left\{\ket{{\bf k}}\right\}$ are multi-local (separable) eigenstates of the operator
$\hat\Sigma_z=\otimes^N_{j=1}\hat\sigma_j^z$ and $\hat{{\mathcal
R}}$ is a multi-local rotation $\hat{{\mathcal
R}}=\otimes^N_{j=1}\hat{R}_j$. With this definition the terms of
the relative entropy of the form $\mathrm{Tr}[\rho_1\log_2\rho_2]$
(we call it here mixed terms) in Eq.~(\ref{GQD}) can be rewritten
as
\begin{equation}
\begin{aligned}
\mathrm{Tr}\left[\rho_T\log_2\hat{\Pi}(\rho_T)\right]&=\mathrm{Tr}\left[\rho_T\log_2\sum_{k} \hat{\mathcal{R}}\ket{{\bf k}}\bra{{\bf k}}\hat{\mathcal{R}}^{\dagger}\rho_T\hat{\mathcal{R}}\ket{{\bf k}}\bra{{\bf k}}\hat{\mathcal{R}}^{\dagger}\right]\\
&=\mathrm{Tr}\left[\rho_T\log_2\hat{\mathcal{R}}\sum_{\bf k}\left( \ket{{\bf k}}\tilde{\rho}_T^{kk}\bra{{\bf k}}\right)\hat{\mathcal{R}}^{\dagger}\right]\\
&=\mathrm{Tr}\left[\tilde{\rho}_T\sum_k\log_2\tilde{\rho}_T^{kk}\ket{{\bf k}}\bra{{\bf k}}\right]\\
&=\sum_k\tilde{\rho}_T^{kk}\log_2\tilde{\rho}_T^{kk}
\end{aligned}
\end{equation}
where we define $\tilde{\rho}_T=\hat{\mathcal{R}}^{\dag}\rho_T\hat{\mathcal{R}}$, $\tilde{\rho}_T^{kk}=\bra{{\bf k}}\hat{\mathcal{R}}^{\dag}\rho_T\hat{\mathcal{R}}\ket{{\bf k}}$, we use that if $\hat{\mathcal{R}}A\hat{\mathcal{R}}^{\dagger}=\tilde{A}$ then $\hat{\mathcal{R}}f\left(A\right)\hat{\mathcal{R}}^{\dagger}=f\left(\tilde{A}\right)$ and the cyclic property of the trace. Using exactly the same line of reasoning for the mixed terms of the relative entropy for the $j$-th qubit we find that
\begin{equation}
\begin{aligned}
\mathrm{Tr}\left[\rho_j\log_2\hat{\Pi}_j(\rho_j)\right]&=\mathrm{Tr}\left[\rho_j\log_2\sum_{l=0}^1 \hat{R}_j\ket{l}\bra{l}\hat{R}_j^{\dagger}\rho_j \hat{R}_j\ket{l}\bra{l} \hat{R}_j^{\dagger}\right]\\&=\sum_{l=0}^1\tilde{\rho}_j^{ll}\log_2\tilde{\rho}_j^{ll}
\end{aligned}
\end{equation}
where $\tilde{\rho}_j^{ll}=\bra{l}\hat{R}_j^{\dag}\rho_j
\hat{R}_j\ket{l}$ and $\ket{l}$ being the two eigenstates of
$\hat\sigma_j^z$. By putting all the terms together and taking
into account the minimization, we obtain the expression for global
discord given in Eq.~(\ref{GDsimp}).


\begin{thebibliography}{99}
\bibitem{Osterloh} A. Osterloh, L. Amico, G. Falci, and R. Fazio, Nature (London) {\bf 416}, 608 (2002).

\bibitem{osborne} T. J. Osborne and M. A. Nielsen, Phys. Rev. A {\bf 66}, 032110 (2002).

\bibitem{Amico} L. Amico,  R. Fazio, A. Osterloh, and V. Vedral, Rev. Mod. Phys. {\bf 80}, 517 (2008).

\bibitem{Ollivier} H. Ollivier and W. H. Zurek, Phys. Rev. Lett. {\bf 88}, 017901 (2001).

\bibitem{Henderson} L. Henderson and V. Vedral, J. Phys. A {\bf 34}, (2001) 6899.

\bibitem{Modi} K. Modi, A. Brodutch, H. Cable, T. Paterek, and V. Vedral, Rev. Mod. Phys. {\bf 84}, 1655 (2012).

\bibitem{varie} R. Dillenschneider, Phys. Rev. B {\bf 78}, 224413 (2008).

\bibitem{giorgi} G. L. Giorgi, B. Bellomo, F. Galve, and R. Zambrini, Phys. Rev. Lett. {\bf 107}, 190501 (2011); G. L. Giorgi, Phys. Rev. A {\bf 84}, 054301 (2011).

\bibitem{werlang}
T. Werlang, C. Trippe, G. A. P. Ribeiro, and G. Rigolin, Phys. Rev. Lett. {\bf 105}, 095702 (2010);
T. Werlang, G. A. P. Ribeiro, and G. Rigolin, Phys. Rev. A {\bf 83}, 062334 (2011).

\bibitem{tomasello}
B. Tomasello, D. Rossini, A. Hamma, and L. Amico, Europhys. Lett. {\bf 96}, 27002 (2011); Int. J. Mod. Phys. B {\bf 26}, 1243002 (2012).

\bibitem{campbell2} S. Campbell, L. Mazzola and M. Paternostro, Int. J. Quant. Inf. {\bf 9}, 1685 (2011).

\bibitem{sarandy1} J. Maziero, L. C. C\'eleri, R. M. Serra and M. S. Sarandy, Phys. Lett. A {\bf 376}, 1540 (2012).

\bibitem{gedik} B. \c{C}akmak, G Karpat, and Z. Gedik, Phys. Lett A {\bf 376}, 2982 (2012).

\bibitem{sarandy2} M. S. Sarandy, T. R. de Oliveira, and L. Amico, Int. J. Mod. Phys. B {\bf 27}, 1345030 (2013).

\bibitem{Simon}
J. Simon,   W. S. Bakr, R. Ma,  M. E. Tai, P. M. Preiss and M. Greiner,
Nature (London) {\bf 472}, 307 (2011).

\bibitem{Friedenauer}
A. Friedenauer, H. Schmitz, J. T. Glueckert, D. Porras, and T. Schaetz, Nat. Phys. {\bf 4}, 757 (2008).

\bibitem{Lanyon}
B. P. Lanyon et al., Science {\bf 334}, 57 (2011).

\bibitem{Britton}
J. W. Britton et al., Nature (London) {\bf 484}, 489 (2012).

\bibitem{Islam}
R. Islam et al, Nat. Comm. {\bf 2}, 377 (2011).

\bibitem{RulliArX11} C. C. Rulli, and M. S. Sarandy, Phys. Rev. A {\bf 84}, 042109 (2011).

\bibitem{Katsura} S. Katsura, Phys. Rev. {\bf 127}, 1508 (1962);
E. Lieb, T. Schultz, and D. Mattis, Annals of Phys. {\bf 16}, 407 (1961);
A. De Pasquale {\it et al.} Eur. Phys. J. Special Topics {\bf 160}, 127 (2008).

\bibitem{Son} W. Son, L. Amico, R. Fazio, A. Hamma, S. Pacazio, and V. Vedral, Europhys. Lett. {\bf 95}, 50001 (2011).

\bibitem{io} D. Girolami, M. Paternostro, and G. Adesso, J. Phys. A: Math. Theor. {\bf 44}, 352002 (2011).

\bibitem{Braga} H.C. Braga, C. C. Rulli, T. R. de Oliveira, and M. S. Sarandy, Phys. Rev. A {\bf 86}, 062106 (2012).

\bibitem{buzek} P. \v{S}telmachovi\v{c} and V. Bu\v{z}ek, Phys. Rev. A {\bf 70}, 032313 (2004).

\bibitem{igloi1} Ferenc Igl\'{o}i and Yu-Cheng Lin, J. Stat. Mech. P06004, (2008); F. Igl\'{o}i and R. Juh\'{a}sz Europhys. Lett. {\bf 81}, 57003, (2008).

\bibitem{campbell} S. Campbell and M. Paternostro, Phys. Rev. A {\bf 82}, (2010) 042324.

\bibitem{gabriele} G. De Chiara, L. Lepori, A. Sanpera, and M. Lewenstein, Phys. Rev. Lett. {\bf 109}, 237208 (2012).

\bibitem{latorre} J. I. Latorre and A. Riera, J. Phys. A: Math. Theor. {\bf 42}, 504002 (2009).

\bibitem{Fisher72}
M.E. Fisher and M.N. Barber, Phys. Rev. Lett. {\bf 28}, 1516
(1972).

\bibitem{Chen} X. Chen, Z.-C. Gu, and X.-G. Wen, Phys. Rev. B {\bf 82}, 155138 (2010).

\bibitem{Smacchia} P. Smacchia, L. Amico, P. Facchi R. Fazio, G. Florio, S. Pascazio, and V. Vedral, Phys. Rev. A {\bf 84}, 022304 (2011).

\bibitem{clusterref} Jian Cui, Luigi Amico, Heng Fan, Mile Gu, Alioscia Hamma and Vlatko Vedral, arXiv:1304.2279.

\bibitem{clusterreview} H. Briegel, D. E. Browne, W. D\"{u}r, R. Raussendorf, M. Van den Nest,  Nature Physics {\bf 5}, 19 (2009).

\bibitem{thermalcluster} R. Raussendorf, S. Bravyi, and J. Harrington, Phys. Rev. A {\bf 71} 062313 (2005);
M. Hadju\v{s}ek, and V. Vedral, New J. Phys. {\bf 12}, 053015 (2010).

\bibitem{commentocluster} The determination of such critical temperature for an $L$-element cluster state is a difficult task~\cite{thermalcluster}.

\bibitem{sonamico} W. Son, L. Amico, F. Plastina, and V. Vedral, Phys. Rev. A {\bf 79}, 022302 (2009).

\bibitem{amicoopt} L. Amico, D. Rossini, A. Hamma, and V. E. Korepin, Phys. Rev. Lett. {\bf 108}, 240503 (2012).

\end{thebibliography}
\end{document}